\documentclass[epj,twocolumn]{webofc}
\usepackage[varg]{txfonts}   
\usepackage{upgreek}
\woctitle{Dark Matter, Hadron Physics and Fusion Physics}
\begin{document}
\title{Dark Forces at DA$\Phi$NE}
%
%

\author{F. Curciarello\inst{1,2}\fnsep\thanks{\email{fcurciarello@unime.it}} on behalf of the KLOE-2 Collaboration
    \thanks{
The KLOE-2 Collaboration:
D.~Babusci,
I.~Balwierz-Pytko,
G.~Bencivenni,
C.~Bloise,
F.~Bossi,
P.~Branchini,
A.~Budano,
L.~Caldeira~Balkest\aa hl,
G.~Capon,
F.~Ceradini,
P.~Ciambrone,
F.~Curciarello,
E.~Czerwi\'nski,
E.~Dan\`e,
V.~De~Leo,
E.~De~Lucia,
G.~De~Robertis,
A.~De~Santis,
P.~De~Simone,
A.~Di~Cicco,
A.~Di~Domenico,
C.~Di~Donato,
R.~Di~Salvo,
D.~Domenici,
O.~Erriquez,
G.~Fanizzi,
A.~Fantini,
G.~Felici,
S.~Fiore,
P.~Franzini,
A.~Gajos,
P.~Gauzzi,
G.~Giardina,
S.~Giovannella,
E.~Graziani,
F.~Happacher,
L.~Heijkenskj\"old,
B.~H\"oistad,
M.~Jacewicz,
T.~Johansson,
K.~Kacprzak,
D.~Kami\'nska,
A.~Kupsc,
J.~Lee-Franzini,
F.~Loddo,
S.~Loffredo,
G.~Mandaglio,
M.~Martemianov,
M.~Martini,
M.~Mascolo,
R.~Messi,
S.~Miscetti,
G.~Morello,
D.~Moricciani,
P.~Moskal,
F.~Nguyen,
A.~Palladino,
A.~Passeri,
V.~Patera,
I.~Prado~Longhi,
A.~Ranieri,
P.~Santangelo,
I.~Sarra,
M.~Schioppa,
B.~Sciascia,
M.~Silarski,
C.~Taccini,
L.~Tortora,
G.~Venanzoni,
W.~Wi\'slicki,
M.~Wolke,
J.~Zdebik}
}

\institute{Dipartimento di Fisica e di Scienze della Terra, Universit\`{a} di Messina
\and
           INFN Sezione Catania}

\abstract{%
The  DA$\Phi$NE $\Phi$-factory is an ideal place to search for forces beyond the Standard Model.
By using the KLOE detector, limits on U-boson coupling $\varepsilon^2$ of the order of  $10^{-5} \div 10^{-7}$ and on the $\alpha_{\rm D} \times \varepsilon^2$ product have been set through the study of the $\Phi$ Dalitz decay, U$\gamma$ events and the Higgsstrahlung process. An improvement of these limits is expected thanks to the KLOE detector and DA$\Phi$NE upgrades of KLOE-2.}

\maketitle
\section{Introduction}
\label{intro}
The Standard Model of physics (SM) represents the most complete theoretical framework currently available to describe fundamental particle composition and interactions. However, there are strong hints of physics beyond it as, for example, non-zero neutrino masses and the discrepancy between calculated and measured value of the muon magnetic moment $a_{\mu}$.
Particularly, the numerous evidences of gravitational anomalies in galaxies and in cosmic microwave background~\cite{PDG} are usually considered as the effect of a large amount of non-barionic matter, called dark matter (DM). 
\begin{figure}[htp!]
\centering
\includegraphics[width=8.5cm,clip]{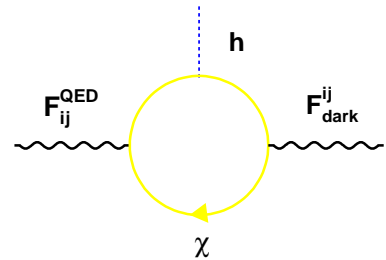}
\caption{Kinetic mixing mechanism; $\rm \chi$ is a dark matter particle, $\rm F^{QED}_{ij}$ and $\rm F^{ij}_{dark}$ are the SM hypercharge and dark tensors respectively, h is the Higgs boson}
\label{fig-1}       
\end{figure}
Many extensions of the SM~\cite{Holdom,U_th1, Fayet,U_th2,U_th6} consider a Weak Interacting Massive Particle (WIMP) as a viable DM candidate and assume that WIMPs are charged under a new kind of interaction called "dark force". This dark force between WIMPs should be mediated by a new gauge vector boson, the U boson, also referred to as dark photon or A', with strength $\alpha_{\rm D}$.
The U is thought to be produced during dark matter annihilation processes and then decay prevalently into leptons if its mass is lower than two proton masses. Usually, the dark photon is associated to an extra abelian gauge symmetry, $U_{\rm D}$, that can couple to SM through a kinetic mixing portal giving rise to a production rate strongly suppressed by the very small coupling~\cite{Holdom,U_th1,U_th2,Fayet,U_th6}.
The ratio of dark and SM hypercharge coupling constants gives the mixing strength $\varepsilon^2$ between the photon and the dark photon~\cite{Holdom}. 
A further hypothesis is that the new symmetry is spontaneously broken by an Higgs-like mechanism, thus resulting in the existence of an additional scalar particle in the dark sector.
A U boson with mass of $\mathcal{O}$(1GeV) and $\varepsilon$ in the range 10$^{-2}$--10$^{-7}$ could
explain all puzzling effects observed in recent astrophysics experiments~\cite{Pamela,AMS,Integral,Atic,Hess,Fermi,Dama/Libra} and account also for the muon magnetic moment anomaly.
For this reason, many efforts have been made in the last years  to find evidence of its existence, with unfortunately null result for the moment~\cite{Mami1,Apex,KLOE_UL1,KLOE_UL2,WASA,HADES,BaBar,a_mu}.

At KLOE-2, dark forces can be probed by using different approaches involving meson decays, $\rm U\gamma$ radiative return processes and Higgsstrahlung.  

In the following we will report on the status of dark forces searches at KLOE/KLOE-2.

\section{DA$\Phi$NE facility and KLOE Detector}
\label{sec-1}
DA$\Phi$NE is an $e^+ e^-$  collider running at the energy $\sqrt{s}=m_\phi=1.0195$~GeV which is located at the National Laboratories of Frascati of INFN. It consists of a linear accelerator, a damping ring, nearly 180~m of transfer lines, two storage rings  that intersect at two points. 

The KLOE detector is made up of a large cylindrical drift chamber (DC), surrounded by a lead scintillating fiber electromagnetic calorimeter (EMC). A superconducting coil around  the EMC provides a 0.52 T magnetic field. The EMC provides measurement of photon  energies, impact point and an accurate measurement of  the time of arrival of particles. The DC is well suited for tracking of the particles and  reaction vertex reconstruction. The calorimeter is divided into a barrel and two end--caps and covers 98\% of the solid angle. The~modules are read out at both ends by 4880 photo--multipliers. Energy and time resolutions are $ \sigma_E /E=~5.7\% /\sqrt{E(\mathrm{GeV})} $ and $ \sigma_t =57 \, \mathrm{ps}~ /\sqrt{E(\mathrm{GeV})}~\oplus 100\, \mathrm{ps}$, respectively. The all-stereo drift chamber, $4 \,m$ in diameter and $ 3.3\, m$ long, has a mechanical structure of carbon fiber-epoxy composite and operates with a light gas mixture (90\% helium, 10\% isobutane). The position resolutions are  $\sigma_{xy} \sim 150\, \upmu \mathrm{m}$ and  $\sigma_z \sim2$~mm. The momentum resolution is $\sigma_{p_\perp} / p_{\perp}$ better than 0.4\%  for large angle tracks. Vertices are reconstructed with a spatial resolution of $\sim 3\, \mathrm{mm}$.
\section{Dark forces at KLOE}
\label{sec-2}
KLOE is particularly suited for dark forces searches thanks to the high data  statistics, the clear event topologies and a good knowledge of backgrounds. Moreover, the cross sections of many processes involving dark photon at $e^+ e^-$ colliders scale with 1/s, compensating the lower luminosity with respect to B-factories. Furthermore, DA$\Phi$NE it's an ideal place to study rare light meson decays. 

At KLOE, dark forces have been investigated by using three different methods: light meson decays,  $\rm U\gamma$ events and by Higgsstrahlung process.

\subsection{$\Phi$ Dalitz decay}
\label{sec-3}
\begin{figure}[htp!]
\centering
\includegraphics[width=5cm,clip]{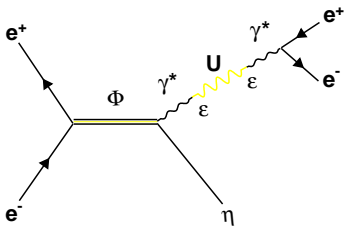}
\caption{U boson production through Dalitz $\phi$ meson decay}
\label{fig-2}       
\end{figure}

The U boson can be produced  in vector (V) to pseudoscalar
(P) meson decays, with a rate that is $\varepsilon^2$ times suppressed with respect to the
ordinary $\mathrm{V} \rightarrow \mathrm{P}$ transitions~\cite{VPU_trans} (see Fig.~\ref{fig-2}). The U boson is supposed to decay into $e^+ e^-$ with a non-negligible branching ratio, thus, $\mathrm{V} \rightarrow\mathrm{ P}$U events are expected to produce a sharp peak in the invariant mass distribution of the electron-positron pair over the
continuum Dalitz background  $\mathrm{V} \rightarrow\mathrm{ P} e^+ e^-$. 

By following this idea, KLOE published two limits on the U-boson coupling $\varepsilon^2$, investigating the $\phi \rightarrow \eta e^+ e^-$ decay, where the $\eta$ meson was tagged by its $\pi^+ \pi^- \pi^0$~\cite{KLOE_UL1} and $3 \pi^0$ decays~\cite{KLOE_UL2}. 
The first analysis selected about 13,000 events by analyzing a data sample of 1.5fb$^{-1}$ integrated luminosity with a 2\% of background contamination. The limit was set by using the Confidence Level Signal (CL$_{\rm S}$) technique~\cite{Feldman,Junk,Read_cls, TLimit}.
This first upper limit (UL) has been then combined, improving sample statistic (1.7fb$^{-1}$) and background rejection, with a new limit derived by tagging the $\eta$ meson by its neutral decay into 3$\pi^0$~\cite{KLOE_UL2}. 
For this new analysis, 30577 events are selected with 3\% background
contamination and an analysis efficiency of 15--30\% at low and high
$e^+e^-$ invariant mass, respectively.
 The determination of the limit is done by varying the $M_{\rm U}$
mass, with 1 MeV step, in the range between 5 and 470 MeV.
Only five bins (5 MeV width) of the reconstructed
$M_{\rm ee}$ variable, centered at $M_{\rm U}$ are considered. For each channel, the irreducible background is
extracted directly from our data after applying a bin-by-bin subtraction of the non-irreducible backgrounds and correcting for the analysis efficiency. The $M_{\rm ee}$
distribution is then fit, excluding the bins used for the upper limit evaluation.

\begin{figure}[htp!]
\centering
\includegraphics[width=6.5cm,clip]{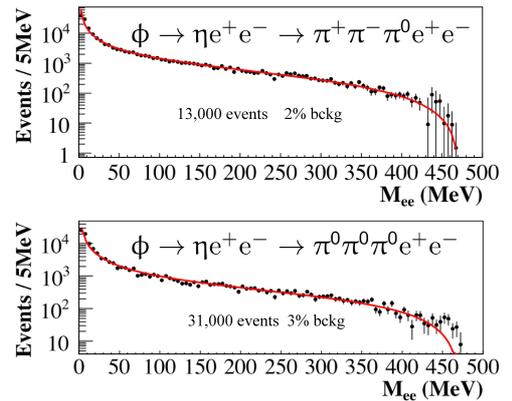}
\caption{$M_{\mathrm{ee}}$ spectrum for the Dalitz decay $\phi \rightarrow \eta e^+ e^-$, with
$\eta \rightarrow \pi^0 \pi^0 \pi^0$ (top) and $\eta \rightarrow \pi^+ \pi^- \pi^0$ (bottom), red line is a fit to the distribution}
\label{fig-3}       
\end{figure}
The limit has been extracted for both $\eta$ decay
channels and then combined. The combined limit evaluation
is done by taking into account the different luminosity, efficiency and relative branching
ratios of the two data samples.
\begin{figure}[htp!]
\centering
\includegraphics[width=6.5cm,clip]{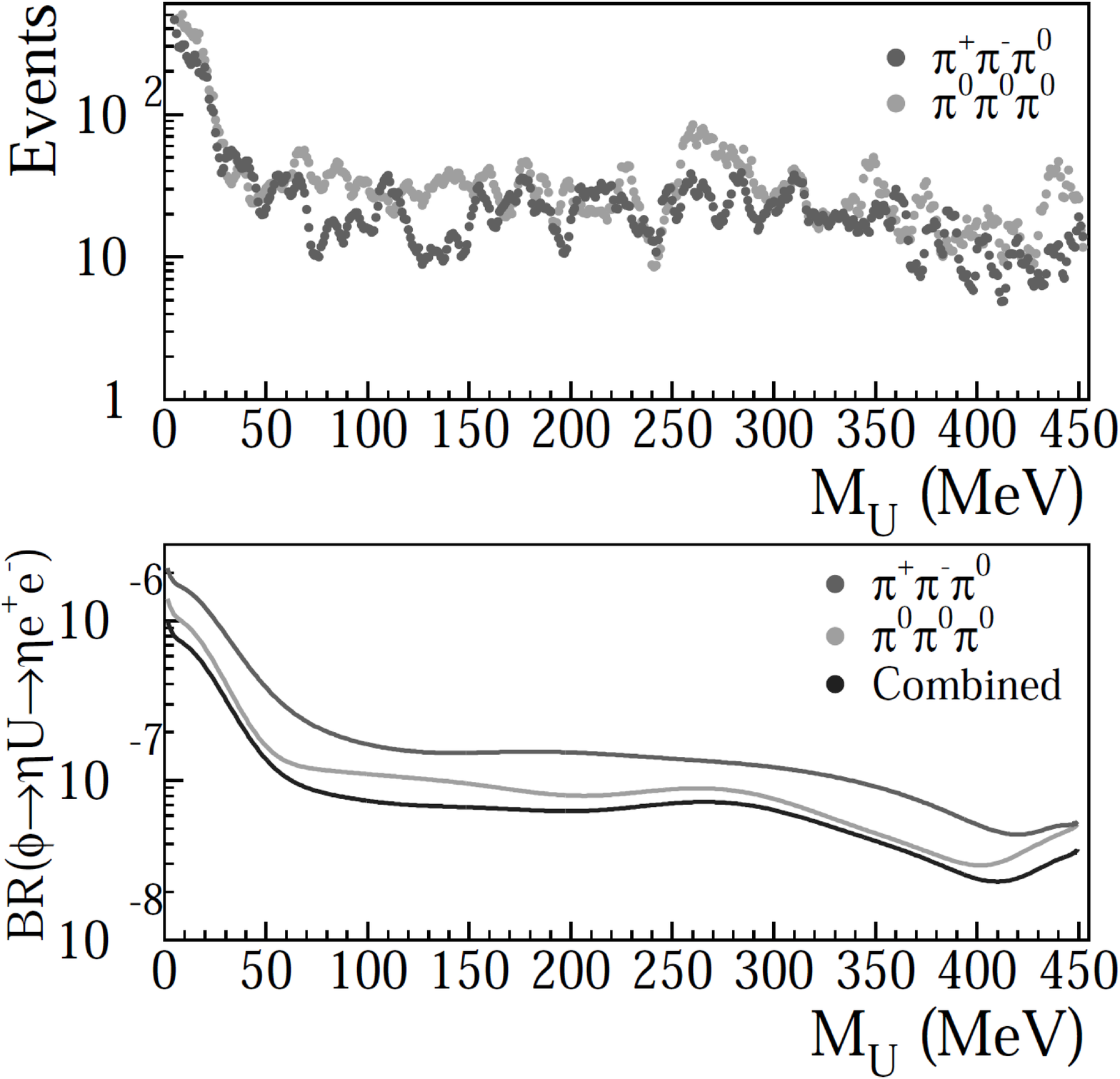}
\caption{Top: upper limit at 90\% CL on the number of events for the decay chain
$\phi \rightarrow \eta \mathrm{U}, \mathrm{U} \rightarrow e^+ e^-$ with $\eta \rightarrow \pi^0 \pi^0 \pi^0$ and $\eta \rightarrow \pi^+ \pi^- \pi^0$. Bottom: smoothed upper limit at 90\% CL on $BR( \phi \rightarrow \eta \mathrm{U}) \times BR(\mathrm{U} \rightarrow e^+ e^-)$ for the
two $\eta $ decay channels and for the combined procedure}
\label{fig-4}       
\end{figure}

In Fig.~\ref{fig-4}, top, the UL at 90\%  confidence level (CL) is shown for each $\eta$ decay channels. In Fig.~\ref{fig-4}, bottom, the smoothed combined upper limit on the branching fraction for the process $\phi \rightarrow \eta \mathrm{U},\mathrm{ U} \rightarrow e^+ e^-$, is compared with evaluations from each of the two $\eta$ decays.
The combined UL on the product
$BR( \phi \rightarrow \eta \mathrm{U}) \times BR(\mathrm{U} \rightarrow e^+ e^-)$ varies from 10$^{-6}$ at small $M_\mathrm{U}$ down to 3 $\times 10^{-8}$.
Using  the Vector Meson Dominance expectation for the transition form factor slope ($b_{\phi\eta}~\sim $1~GeV~$^2$) an UL on the parameter $\varepsilon^2$ at 90\% CL can be derived of $ \varepsilon^2 < 1.7 \times 10^{-5 }$ for $30 < M_\mathrm{U} < 400$~MeV, and  for the sub-region $50 < M_\mathrm{U} < 210$~MeV of $ \varepsilon^2 < 8.0 \times 10^{-6}$.
This limit~\cite{KLOE_UL2} excludes wide range of U-boson parameters as possible explanation of the $a_{\mu}$ discrepancy.
\subsection{$\rm U \gamma$ events}
\label{sec-4}
\begin{figure}[htp!]
\centering
\includegraphics[width=5cm,clip]{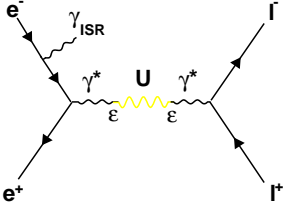}
\caption{Resonant U-boson production in $e^+\mathrm{ e}^-$ collisions with Initial State Radiation (ISR) emission}
\label{fig-5}       
\end{figure}

The U boson can be produced in any process in which a photon is involved but with a rate strongly suppressed by the small coupling. At KLOE, an interesting channel is the U-boson radiative production
 $e^+ e^- \rightarrow \mathrm{U} \gamma, \, \mathrm{U} \rightarrow l^+ l^-\,, l=e,\mu$ shown in Fig.~\ref{fig-5}.

It is considered a very simple and clean channel, independent of the existence and of the details of the Higgs sector of the secluded group. The expected signal is a Breit-Wigner peak in the invariant mass distribution of the lepton pair induced by the mechanism of photon radiative return and corresponding to U-boson resonant production~\cite{babayaga_article}. 

KLOE investigated both the $\mu^+ \mu^- \gamma$ and $e^+ e^- \gamma$ final states.

\subsubsection{$\mu^+\mu^-\gamma$ limit}
\label{sec-4.1}
This search employed a data sample collected in 2002 at DA$\Phi$NE \, $e^+ e^-$  collider with an integrated luminosity of 239.3 pb$^{-1}$. 
The U-boson peak would appear in the dimuon mass spectrum. 
The $\mu^+ \mu^- \gamma$ event selection requires two tracks of opposite charge with $50^\circ\!<\!\theta\!<\!130^\circ$ and an undetected photon whose momentum
points at small polar angle ($\theta\!<\!15^\circ,\ \!>\!165^\circ$)~\cite{KLOE_U,KLOE_pi_FF}.  Pions and muons are separated by means of the variable $M_{\rm trk}$ defined as the mass of particles $x^+,\ x^-$ in the $e^+ e^- \to x^+ x^- \gamma$ process, calculated from the overdetermined system of kinematical constraints of the reaction. 
The $M_{\rm trk}$ values between  80--115 identify muons while $M_{\rm trk}$ values $>$130 MeV identify pions. To improve $\pi/\mu$ separation a cut based on the quality of fitted tracks has been used resulting in a suppression of the left tail of $\pi\pi\gamma$ $M_{\rm trk}$ distribution up to 40\%~\cite{KLOE_U}.
At the end of the analysis  chain, the residual background is obtained by fitting the  $M_{\rm trk}$ data distribution with Monte Carlo (MC) simulations describing signal, $\pi^+ \pi^- \gamma$ and $\pi^+ \pi^- \pi^0$ backgrounds plus a distribution obtained from data for the $e^+e^-\gamma$ ~\cite{KLOE_U,KLOE_pi_FF}.
\begin{figure}[htp!]
\centering
\includegraphics[width=7cm,clip]{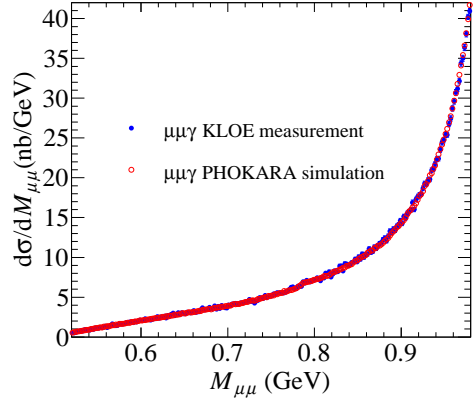}
\caption{ $\mu^+ \mu^- \gamma$ absolute cross section for data compared with the QED NLO MC PHOKHARA prediction in the 520--980~MeV energy range}
\label{fig-6}       
\end{figure}

\begin{figure}[htp!]
\centering
\includegraphics[width=7cm,clip]{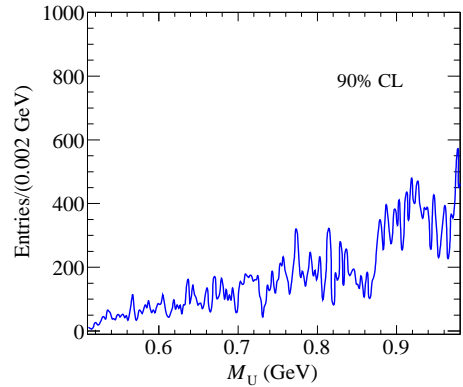}
\caption{Upper limit on number of U-boson events excluded at the 90\% CL}
\label{fig-7}       
\end{figure}
Finally, we derived the differential cross section $\rm d\sigma_{\mu \mu \gamma}/\rm  dM_{\mu\mu}$ achieving an excellent agreement between the measurement and the simulation based on PHOKHARA~\cite{PHOKHARA}, as shown in Fig.~\ref{fig-6}. No structures are visible in the $M_{\mu\mu}$ spectrum.
We extracted the limit on the number of U-boson candidates
through the CL$_{\rm S}$ technique~\cite{Feldman,Junk,Read_cls, TLimit} by comparing the expected and observed $\mu^+\mu^- \gamma$ yield, and a MC generation of the U-boson signal which takes into account the $M_{\mu\mu}$ invariant mass resolution (1.5~MeV to 1.8~MeV as $M_{\mu\mu}$ increases), see Fig.~\ref{fig-7}. A systematic error of 1.4--1.8\% on the expected background has also been taken into account in the limit evaluation.
The limit on the number of U-boson events has been converted in terms of the kinetic mixing parameter $\varepsilon^2$ by using the following formula:
\begin{equation}
 \varepsilon^2(M_{ll})= \frac{N_{\rm{CLS}}(M_{ll})/ (\epsilon_{\rm{eff}}(M_{ll}) \times L(M_{ll}))}{H(M_{ll}) \times I(M_{ll})},
\label{eq.2}
\end{equation}
where l=$\mu$, $\epsilon_{\rm{eff}}(M_{ll})$ represents
the overall efficiency (1-15\% as $M_{\mu\mu}$ increases~\cite{KLOE_U}), $L(M_{ll})$ is the integrated luminosity, $H(M_{ll})$ is the radiator function obtained from QED including NLO corrections~\cite{KLOE_U}, and $I(M_{ll})$ is the effective U cross section~\cite{Fayet}.
The limit on $\varepsilon^2$~\cite{KLOE_U} is of 1.6~$\times 10^{-5}$ and 8.6 $\times 10^{-7}$ in the 520--980 MeV energy range. This limit represents the first one derived by a direct study of the $\mu^+\mu^-\gamma$ channel.

\subsubsection{$e^+ e^-\gamma$ limit}
\label{sec-4.2}
\begin{figure}[htp!]
\centering
\includegraphics[width=7.7cm,clip]{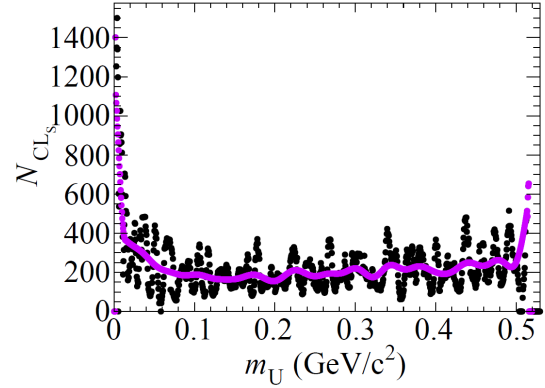}
\caption{Upper limit on number of U-boson candidates at 90\% CL, violet line is a smoothing}
\label{fig-9}       
\end{figure}

\begin{figure*}[htp!]
\centering
\includegraphics[width=16.7cm,clip]{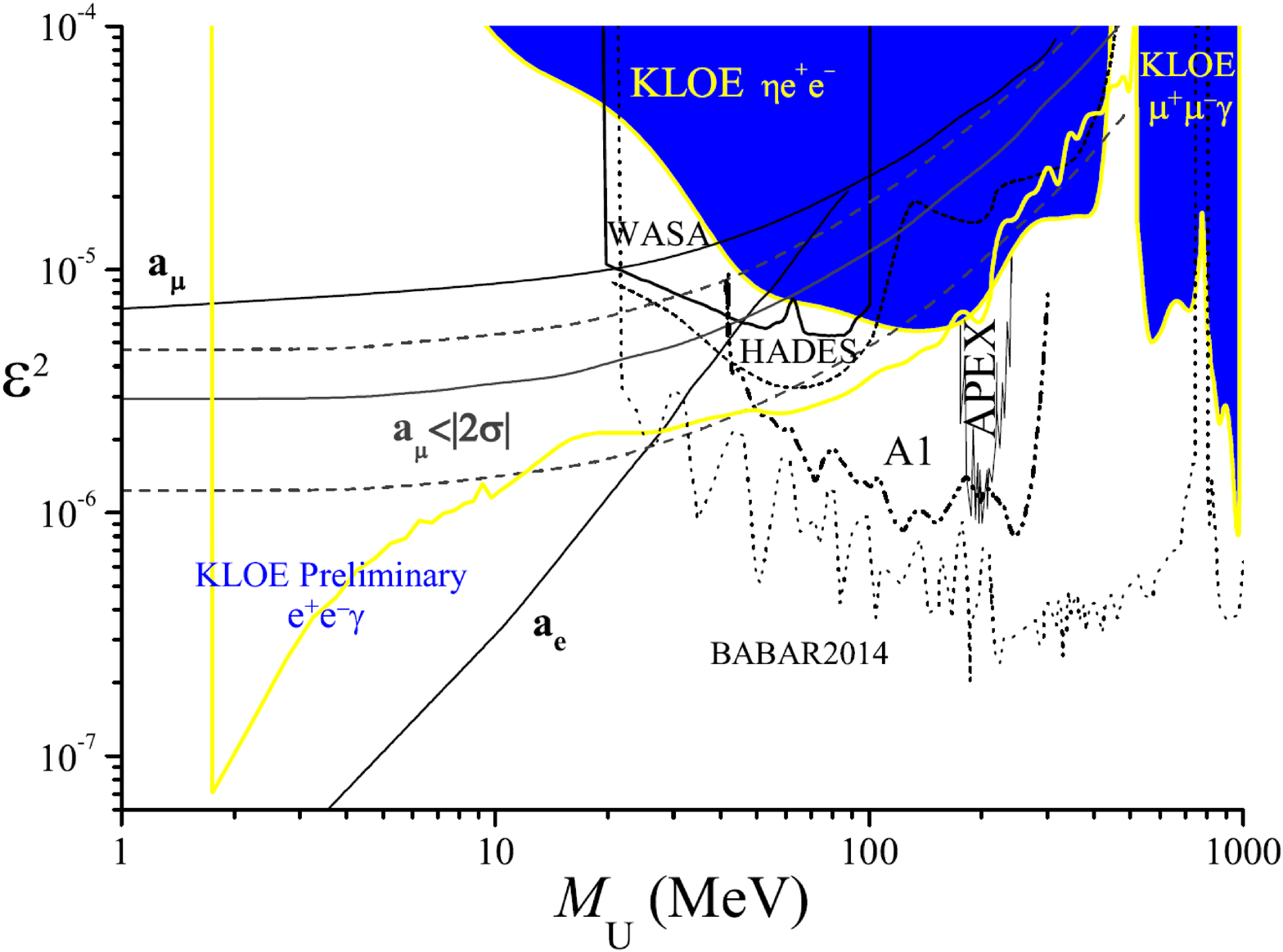}
\caption{90\% CL  KLOE exclusion plots for $\varepsilon^2$ as a function of the U-boson mass~\cite{KLOE_UL1, KLOE_UL2,KLOE_U}. Limits from  A1~\cite{Mami1}, Apex~\cite{Apex}, WASA~\cite{WASA}, HADES~\cite{HADES} and BaBar ~\cite{BaBar} are also shown. The black and grey lines are the limits from the muon and electron magnetic moment anomalies~\cite{a_mu}}
\label{fig-11}       
\end{figure*}

The KLOE search for the $\mathrm{U} \to e^+ e^-$ employed a data sample of an integrated luminosity of 1.54~fb$^{-1}$. In this analysis the hard ISR photon has been explicitly detected in the calorimeter barrel by requiring for the charged leptons and photon a polar angle 50$^{\circ}$< $\theta$ <130$^{\circ}$. This large-angle event selection allowed us to have a sufficient statistics to reach the dielectron mass threshold. A cut on the track mass variable ($ \rm -70~MeV < M_{\rm trk}< \rm 70~MeV$) was again applied to remove background contamination from $\mu^+ \mu^- \gamma$, $\pi^+\pi^- \gamma$, $\pi^+ \pi^- \pi^0$, $e^+ e^- \to \gamma \gamma$ (where the photon converts into an $e^+ e^-$ pair) and other $\phi$ decays. At the end of the analysis chain the background contamination is less than 1.5\%. 
\begin{figure}[htp!]
\centering
\includegraphics[width=7.7cm,clip]{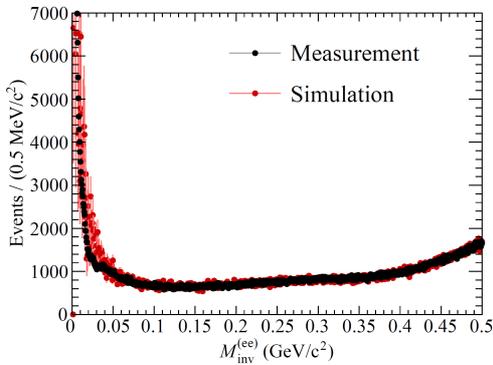}
\caption{Dielectron invariant mass distribution compared with Babayaga-NLO simulation}
\label{fig-8}       
\end{figure}
Figure~\ref{fig-8} shows the final  dielectron invariant-mass distribution demonstrating excellent
agreement with a Babayaga-NLO MC simulation~\cite{babayaga_article}.
Also in this case no resonant U-boson peak was observed and again the CL$_{\rm S}$ technique is applied to estimate N$_{CLs}$ , the
number of U-boson signal events excluded at 90\% CL (see Fig.~\ref{fig-9}). 
A preliminary limit  on the kinetic mixing parameter as a function
of $\rm M_{\rm U}$ was set by using eq.~\ref{eq.2}  with $l=e$, $\epsilon_{\rm eff}(M_{ll})$=1.5–2.5\% and L=1.54 fb$^{-1}$.
The resulting exclusion plot is shown in Fig.~\ref{fig-11} with all other existing limits in the region of 0--1000~MeV. 

\subsection{Dark Higgsstrahlung process}
\label{sec-5}
A natural consequence of the U-boson existence is the breaking of the $\rm U_D$ symmetry associated to it by a Higgs-like mechanism through an additional scalar particle,  the dark Higgs $\mathrm{h}^{\prime}$. The expected signature depends on the  U and $\mathrm{h}^{\prime}$ mass hierarchy . If the $\mathrm{h}^{\prime}$ is lighter than the U boson it turns out to be very long-lived and escapes detection (see Ref.~\cite{Batell}).  The expected signal will be a lepton pair from the U-boson decay plus missing energy. 
\begin{figure}[htp!]
\centering
\includegraphics[width=5cm,clip]{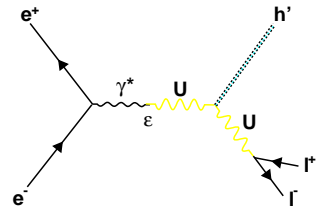}
\caption{Diagram of the U-boson production through Higgsstrahalung process}
\label{fig-10}       
\end{figure}

\begin{figure}[htp!]
\centering
\includegraphics[width=6.2cm,clip]{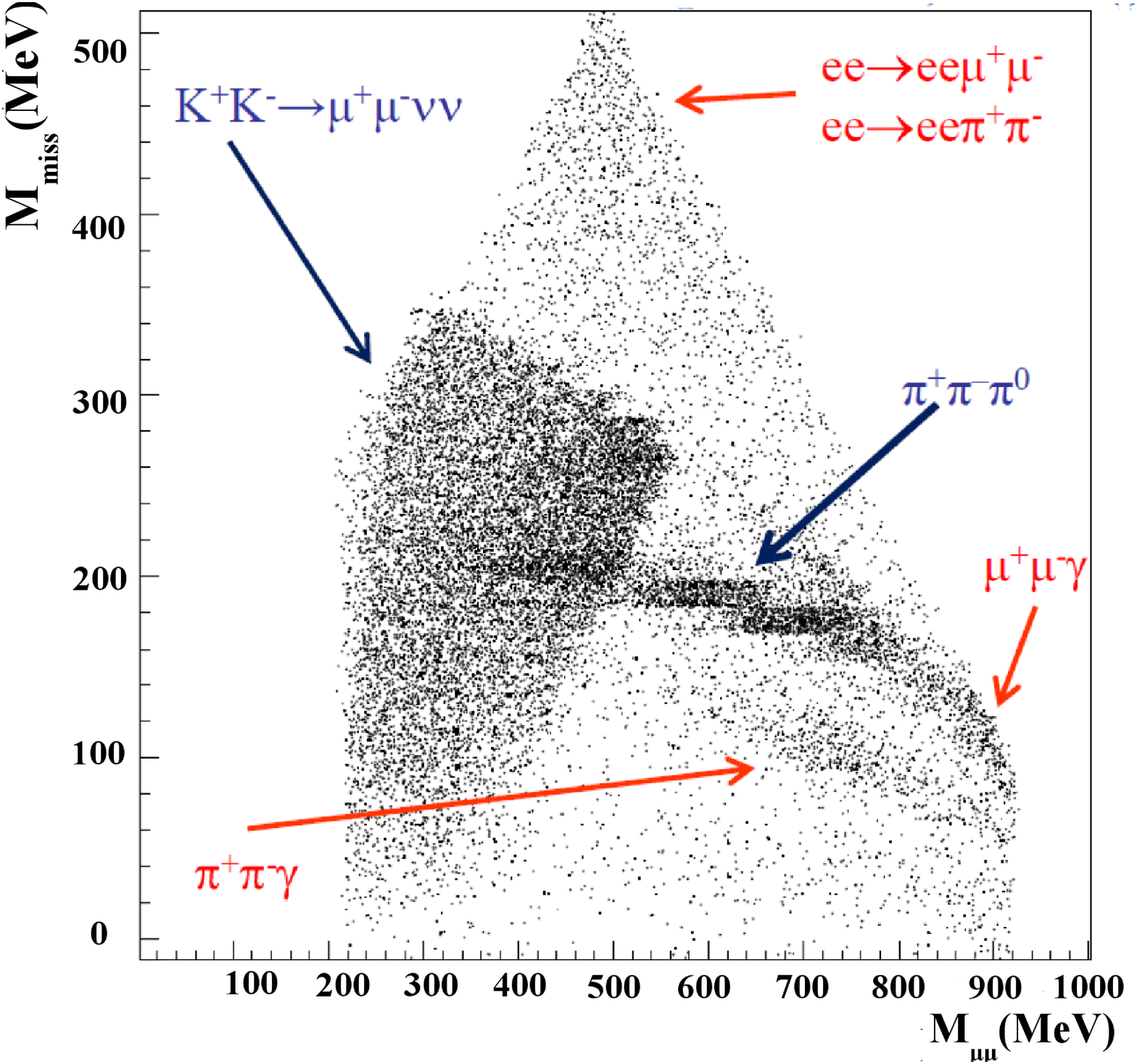}
\includegraphics[width=6.2cm,clip]{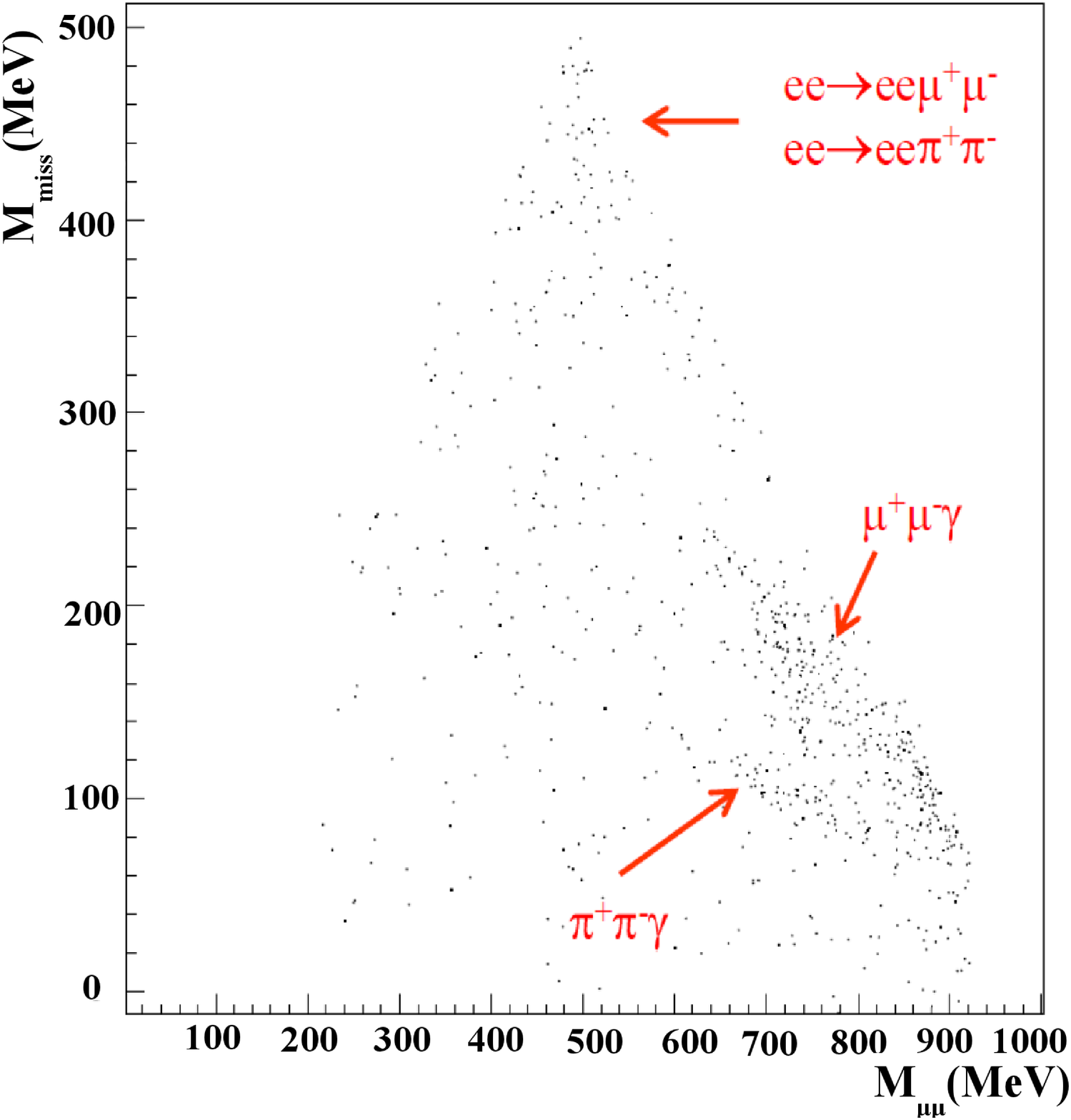}
\caption{Top panel: $M_{\mathrm{miss}}$ Vs $M_{\mu\mu}$  for on-peak sample (1.65~fb$^{-1}$ integrated luminosity). Bottom panel:  $M_{\mathrm{miss}}$ Vs $M_{\mu\mu}$ for the off-peak sample, (0.2~fb$^{-1}$ integrated luminosity)}
\label{fig-12}       
\end{figure}
This intriguing hypothesis (see Fig.~\ref{fig-10}), can be observed at KLOE if $m_\mathrm{U}+m_{\mathrm{h}^{\prime}} < m_{\phi}$. 
\begin{figure}[htp!]\vspace{-0.5cm}
\centering
\includegraphics[width=6.5cm,clip]{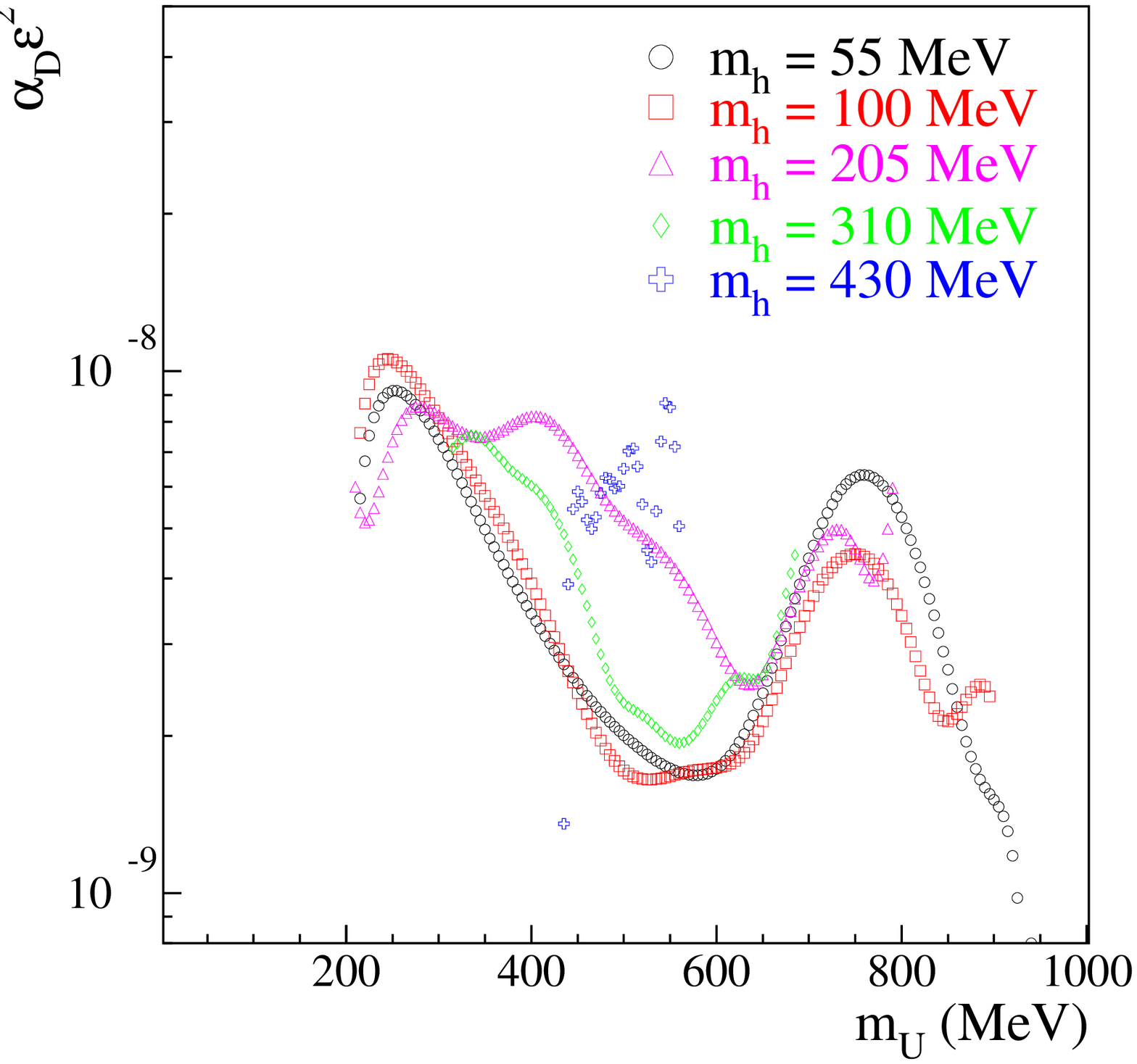}\vspace{-0.6cm}
\includegraphics[width=6.5cm,clip]{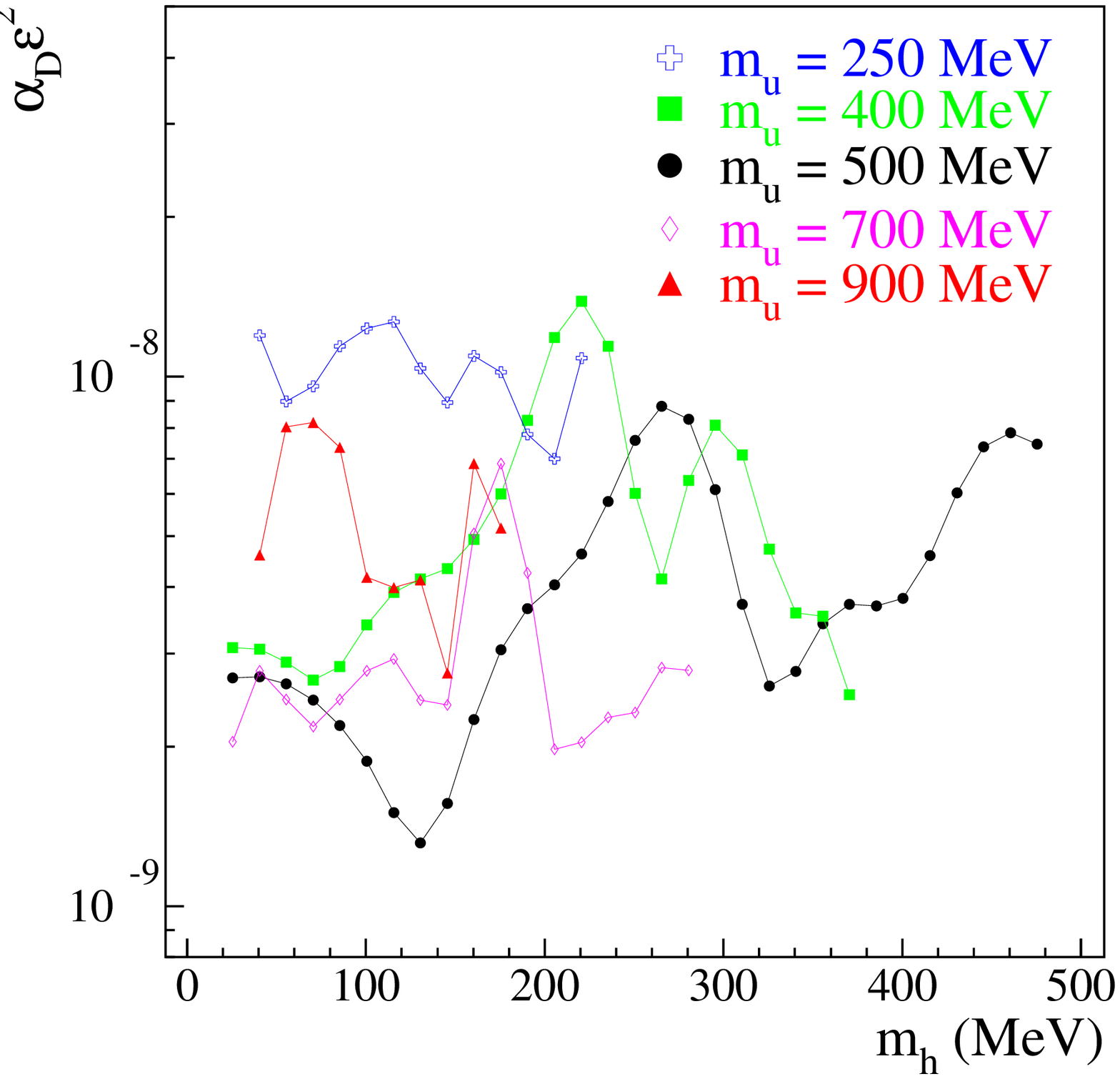}
\caption{Preliminary combined 90\% CL upper limits in $\alpha_\mathrm{D} \times \varepsilon^2$ as a function of $M_{\mu\mu}$ for
different values of $m_{\mathrm{h}^{\prime}}$ (upper plot) and as a function of $M_{\mathrm{miss}}$ for different values
of $M_\mathrm{U}$ (bottom plot plot)}
\label{fig-13}       
\end{figure}
The production cross section of the dark Higgsstrahlung process is proportional
to the product $\alpha_\mathrm{D} \times \varepsilon^2$ and depends on the boson masses~\cite{Batell}. 
Moreover, compared to the B-factory case~\cite{dark_higgs_1}, KLOE benefits of the 1/s factor and of the resonance-like behavior expected for the production cross section~\cite{Batell}.
KLOE is studying the Higgsstrahlung process in the energy range $2 m_{\mu}<M_\mathrm{U} < 1000$~MeV, by limiting its search to the $e^+ e^- \rightarrow \mathrm{\mathrm{h}^{\prime}\mathrm{U}}, \mathrm{U} \rightarrow \mu^+ \mu^- $, $\mathrm{h}^{\prime}$ invisible~\cite{Enrico}.
 The analysis has been performed by using a data sample of 1.65~fb$^{-1}$ collected at center of mass energy $E_{\mathrm{cm}}$ corresponding to the $\phi$ peak ($\sim$~1019~MeV) and a data sample of 0.2~fb$^{-1}$ at $E_{\mathrm{cm}}=1000$~MeV (off-peak sample).
The event selection requires~\cite{Enrico} events with only two opposite charge tracks, with a reconstructed vertex inside a $4 \times 30$~cm cylinder around the interaction point (IP).  Each track must have an associated EMC cluster and the visible momentum direction has to be in the barrel ($|\cos \theta| < 0.75$).
After the missing energy and the particle identification selections, a huge background from
$\phi \rightarrow \mathrm{K}^+ \mathrm{K}^-, \mathrm{K}^{\pm} \rightarrow\mu^{\pm} \nu$ events survives in the on-peak sample. A cut based on the vertex-IP distance and on the goodness of the track fitting $\chi^2$ allowed to reduce the $\mathrm{K}^{\pm}$ background. 
The expected signal would appear in the $M_{\mu\mu}-M_{\mathrm{miss}}$ bi-dimensional spectra. In order too keep most of the
signal in one bin only, a 5~MeV bin width in the $M_{\mu\mu}$ has been chosen while for $M_{\mathrm{miss}}$ a variable binning of 15, 30 and 50~MeV widths is used. Results are shown in Fig.~\ref{fig-12} for the on-peak and off-peak sample. In the second case of course the huge backgrounds coming from $\phi$ resonant processes are suppressed.
No signal signature has been observed and a bayesian limit on the number of signal events at 90\% CL has been evaluated, bin-by-bin, for the on-peak and off-peak sample separately. A conservative systematic error of about 10\% is considered for the limit extraction. Results have been translated in terms of $\alpha_\mathrm{D} \times \varepsilon^2$ by using the integrated luminosity information, the signal efficiency (15--25\%), the dark Higgsstrahlung cross section and the branching fraction of the $\mathrm{U} \rightarrow \mu^+ \mu^-$ decay~\cite{Batell}.
Results were then combined by taking into account the different integrated luminosities, different signal efficiencies and cross sections of the two data samples. The combined preliminary upper limits projected in the $M_{\mu\mu}$ and $M_{\mathrm{miss}}$ directions and slightly smoothed are shown in Fig.~\ref{fig-13}. Values of the order of 10$^{-9} \div 10^{-8}$ in $\alpha_\mathrm{D} \times \varepsilon^2$ are excluded at 90\% CL for a large range of the dark photon and dark Higgs masses. These limits translate in $\varepsilon \sim 10^{-3} -10^{-4}$ for $\alpha_\mathrm{D}=\alpha_{\mathrm{em}}$ and are in agreement and complementary with BaBar results~\cite{dark_higgs_1} as they refer to the same process in a different final state and phase space region.

\newpage
\section*{Conclusions}

The KLOE2 Collaboration searched for dark forces by performing five analyses on four different processes.
No evidence for the existence of new "dark" particles was found. 90\% CL limits of $10^{-7}-10^{-5}$ have been set on the kinetic mixing parameter $\varepsilon^2$ in the energy range 5--980 MeV. Bayesian 90\% CL limits on the product  $\alpha_{\rm D} \times \varepsilon^2 $ of $10^{-9}-10^{-8}$ in the parameter space 2$m_{\mu}$ < $M_{\rm U}$ < 1000 MeV and 10 < m$_{\rm h'}$ < 500 MeV have been extracted. DA$\Phi$NE luminosity and detectors upgrades, especially the insertion of the inner tracker,  are expected to improve these limits of a factor of about two in the next KLOE-2 experiment.
%

\begin{thebibliography}{50}
%
%
\bibitem{PDG} Particle Data Group Collaboration, J. Beringer et al.,  
Phys. Rev. D \textbf{86},  010001 (2012)
%
\bibitem{Holdom} B. Holdom,  Phys.\ Lett.\ B \textbf{166}, 196 (1985)
%
\bibitem{U_th1} C. Boehm, P. Fayet, Nucl. Phys.\ B \textbf{683}, 219 (2004) 
%
\bibitem{Fayet}P. Fayet, Phys.\ Rev.\ D \textbf{75}, 115017 (2007)  
%
\bibitem{U_th2}  Y. Mambrini, J. Cosmol.\ Astropart.\ Phys.\ \textbf{1009}, 022 (2010) 
%
\bibitem{U_th6} M. Pospelov, A. Ritz, M.B. Voloshin, Phys.\ Lett.\ B \textbf{662}, 53 (2008)
%
\bibitem{Pamela} O. Adriani, et al., Nature \textbf{458}, 607 (2009)
%
\bibitem{AMS} M. Aguilar, et al. Phys.\ Rev.\ Lett.\ \textbf{110}, 141102 (2013) 
%
\bibitem{Integral} P. Jean, et al., Astronomy Astrophysics \textbf{407}, L55 (2003) 
%
\bibitem{Atic} J. Chang, et al., Nature \textbf{456}, 362 (2008) 
%
\bibitem{Hess} F. Aharonian, et al., Phys.\ Rev.\ Lett.\ \textbf{101}, 261104 (2008) 
%
\bibitem{Fermi} A. A. Abdo, et al., Phys.\ Rev.\ Lett.\ \textbf{102}, 181101 (2009) 
%
\bibitem{Dama/Libra} R. Barnabei, et al., Eur.\ Phys.\ J.\  C \textbf{56}, 333 (2008) 

\bibitem{Mami1}H. Merkel, et al., Phys.\ Rev.\ Lett.\ \textbf{112} 221802, (2014) 
%
\bibitem{Apex} S. Abrahamyan, et al.,  Phys.\ Rev.\ Lett.\ \textbf{107}, 191804 (2011) 
%
\bibitem{KLOE_UL1}F. Archilli, et al., Phys.\ Lett.\ B \textbf{706}, 251 (2012) 
%
\bibitem{KLOE_UL2}D. Babusci, et al., Phys.\ Lett.\ B \textbf{720}, 111 (2013)
%
\bibitem{WASA} P. Adlarson, et al., Phys.\ Lett.\ B \textbf{726}, 187 (2013) 
%
\bibitem{HADES} G. Agakishiev et al., Phys.\ Lett.\ B \textbf{731}, 265 (2014) 
%
\bibitem{BaBar}  B. Aubert, et al., Phys.\ Rev.\ Lett.\ \textbf{103}, 081803 (2009); J. D. Bjorken, R. Essig, P. Schuster, and N. Toro, Phys. Rev. D \textbf{80}, 075018 (2009); J. P. Lees et al., Phys.\ Rev.\ Lett.\ \textbf{113}, 201801 (2014) 
\bibitem{a_mu} M. Pospelov, Phys.\ Rev.\ D \textbf{80},  095002 (2009) 
%
\bibitem{VPU_trans}M. Reece, L.T. Wang, JHEP \textbf{07}, 051 (2009)


%
\bibitem{Feldman} G. J. Feldman and R. D. Cousins, Phys. Rev. D \textbf{57}, 3873 (1998) 

\bibitem{Junk} T. Junk,  Nucl. Instr. Meth. A \textbf{434}, 435 (1999)  

\bibitem{Read_cls}A. L. Read, J. Phys. G: Nucl. Part. Phys. \textbf{28}, 2693 (2002) 

\bibitem{TLimit} http://root.cern.ch/root/html/TLimit.html
%
\bibitem{babayaga_article} L. Barz\`{e} et al., Eur. Phys. J. C \textbf{71}, 1680 (2011)
%
\bibitem{KLOE_U} D. Babusci et al., Phys. Lett. B \textbf{736}, (2014) 459
%
\bibitem{KLOE_pi_FF}D. Babusci, et al., Phys. Lett. B \textbf{720}, (2013) 336
%
\bibitem{PHOKHARA} H. Czy\.{z} , A. Grzelinska, J.H. K\"uhn, G. Rodrigo, Eur. Phys. J. C \textbf{39}, (2005) 411
%
\bibitem{Batell}B. Batell et al., Phys. Rev. D \textbf{79}, 11508 (2009)
%
\bibitem{Enrico} D. Babusci et al., arXiv:1501.06795 [hep-ex]
  	
%
\bibitem{dark_higgs_1}J.P. Lees et al. (BaBar Collab.) Phys. Rev. Lett. \textbf{108}, 211801 (2012)
%
\end{thebibliography}
%
%

\end{document}